\documentclass[12pt,draftcls,onecolumn]{IEEEtran}
\usepackage{latexsym}
\usepackage{graphicx}
\usepackage{array}
\usepackage{amsmath}
\usepackage{amsfonts}
\usepackage{amssymb}

\newtheorem{thm}{Theorem}

\newtheorem{prop}{Proposition}

\begin{document}

\title{Compound Outage Probability and Capacity of a Class of Fading MIMO Channels with Channel Distribution Uncertainty}

\author{Ioanna Ioannou,  Charalambos D. Charalambous and Sergey Loyka

\thanks{Manuscript received XXX}

\thanks{I. Ioannou and C.D. Charalambous are with the ECE Department, University of Cyprus, 75 Kallipoleos Avenue, P.O. Box 20537, Nicosia, 1678, Cyprus, e-mail: aioannak@yahoo.gr, chadcha@ucy.ac.cy}

\thanks{S. Loyka is with the School of Information Technology and Engineering, University of Ottawa, Ontario, Canada, K1N 6N5, e-mail: sergey.loyka@ieee.org}
}

\maketitle

\begin{abstract}
Outage probability and capacity of a class of block-fading MIMO channels are considered with partial channel distribution information. Specifically, the channel or its distribution are not known but the latter is known to belong to a class of distributions where each member is within a certain distance (uncertainty) from a nominal distribution. Relative entropy is used as a measure of distance between distributions. Compound outage probability defined as min (over the transmit signal distribution) -max (over the channel distribution class) outage probability is introduced and investigated. This generalizes the standard outage probability to the case of partial channel distribution information. Compound outage probability characterization (via one-dimensional convex optimization), its properties and approximations are given. It is shown to have two-regime behavior: when the nominal outage probability decreases (e.g. by increasing the SNR), the compound outage first decreases linearly down to a certain threshold (related to relative entropy distance) and then only logarithmically (i.e. very slowly), so that no significant further decrease is possible. The compound outage depends on the relative entropy distance and the nominal outage only, all other details (nominal fading and noise distributions) being irrelevant. The transmit signal distribution optimized for the nominal channel distribution is shown to be also optimal for the whole class of distributions. The effect of swapping the distributions in relative entropy is investigated and an error floor effect is established. The compound outage probability under $L_p$ distance constraint is also investigated. The obtained results hold for a generic channel model (arbitrary nominal fading and noise distributions).
\end{abstract}


\section{Introduction}
\IEEEPARstart{M}{ultiple}-input multiple-output (MIMO) wireless systems have received significant attention due to the promise of high spectral efficiency \cite{Telatar}\cite{Foshini}, which has been extensively investigated. As with any wireless system, their channel capacity depends significantly on the channel state information available at the transmitter and the receiver as well as the fading statistics experienced by the channel \cite{Biglieri}. When the fading process is egrodic (i.e. the channel "reveals" its statistics to a single codeword), an appropriate performance indicator is ergodic capacity \cite{Telatar}\cite{Biglieri}. On the other hand, when the channel is block-fading (or quasi-static), i.e. stays fixed during a codeword transmission and changes from codeword to codeword, its Shannon capacity is zero in many cases of practical interest (i.e. Rayleigh fading) so that outage capacity (capacity versus outage) and outage probability (for given target rate) are appropriate performance indicators \cite{Telatar}-\cite{Tse}. In the block-fading (quasi-static) regime, the channel capacity is not affected by the receiver channel state information \footnote{since the receiver can always learn the channel via a training sequence, which results in asymptotically-negligible loss in the capacity in the quasi-static mode \cite{Biglieri}, full CSI at the receiver can be assumed that significantly simplifies the analysis.} but depends significantly on the channel state information available at the transmitter  \cite{Biglieri}\cite{Goldsmith}\cite{Caire}. Since the channel state information is obtained via channel measurements, its accuracy may be limited due to variability and difficult propagation conditions (e.g. low SNR) in a wireless channel. The channel state information at the transmitter is further limited due to limitations of the feedback channel (if any). This situation can be modeled via a compound channel model, where the true channel is not known but it is known to belong to a certain (limited) class of channels and the corresponding compound channel capacity theorems have been established \cite{Dobrushin}-\cite{Lapidoth}. While these theorems treat all channels in the class equally and build a code that performs well on any such channel, the corresponding capacity is typically limited by the worst channel in the class and may be low, even though most channels in class are good and the worst channel is realized with low probability, i.e. it is a conservative performance indicator. To avoid this problem, a concept of composite channel has been introduced \cite{Lapidoth}\cite{Effros}, where each channel in a class has associated probability measure, so that bad low-probability channels do not penalize significantly the performance metric. The corresponding channel capacity theorems can be proved via the concept of information density \cite{Verdu}\cite{Effros} or using the compound channel approach \cite{Biglieri}\cite{Caire}.

Another possibility to model the uncertainty of channel state information is to assume that the transmitter knows only the channel distribution but not the channel itself. A number of results on MIMO channel capacity have been obtained under this assumption \cite{Goldsmith}-\cite{Caire}. A comprehensive review of the impact of channel uncertainty on its performance and corresponding coding/decoding strategies can be found in \cite{Lapidoth}. A concise review of more recent activities on MIMO channels is available in \cite{Wiesel}. The compound MIMO channel capacities under the trace and spectral norm constraints have been studied recently in \cite{Wiesel}-\cite{Denic}. A construction of a code approaching the compound channel capacity can be found in \cite{Kose}.

In this paper, we consider a situation where even the channel distribution information is not available at the transmitter; rather, the transmitter knows that the channel distribution belongs to a certain class centered around a nominal distribution. This models a practical scenario where the channel distribution information is obtained from multiple but limited measurements, so that the true distribution is known only with finite accuracy (typically related to the number of independent samples used for estimation). This also models a dynamic scenario where the channel distribution information obtained from past measurements may be outdated. The uncertainty in the channel distribution information may also be related to the limitation of the feedback channel used to supply this information to the transmitter. We assume a quasi-static (block-fading) scenario so that channel state information at the receiver is irrelevant. Our channel model is quite generic: we do not assume any particular nominal channel distribution and even the channel noise can be arbitrary (except for examples, where particular distribution and noise are considered), so that the results are general too. Relative entropy between two distributions is used as a measure of distance, so that the distribution uncertainty class includes all distributions within certain relative entropy distance of the nominal one. Similar approach was adopted in \cite{Charalambous} to study the ergodic capacity under channel distribution uncertainty \footnote{while the impact of channel distribution on the capacity is quite mild in the ergodic regime (due to averaging over the channel statistics), it is much stronger in the non-ergodic regime (no averaging) \cite{Tse}.} and in \cite{Charalambous 2} to investigate an optimal control of stochastic uncertain systems. A justification of relative entropy as a measure of distance between distributions can be found in e.g. \cite{Cover}\cite{Verdu-2}. Our results on compound outage probability provide further justification, as they indicate that the relative entropy distance limits the achievable outage probability (capacity) via the error floor effect. When the nominal outage probability is negligible and the distance is small, the compound outage probability equals to the relative entropy distance (regardless of all other details).

Since the channel is block-fading, the outage probability and capacity are considered as main performance metrics, which we term "compound outage probability/capacity" to emphasize that it applies to a class of fading distributions (i.e. "compound distribution") rather than any particular one. This parallels the concept of compound channel, where a code is designed to operate on any member in the class. In our case, a code is designed to operate for any channel distribution in the class, so that the compound outage probability involves maximization over all feasible channel distributions and minimization over the transmitted signal distribution (subject to the power constraint), and the corresponding compound outage capacity is derived from it. We also consider a scenario where the transmitted signal distribution is fixed a priori (e.g. universal code design).

The system/channel model and the performance metrics (outage probability and capacity) are introduced in Section \ref{sec:System model}. Compound outage probability is defined and investigated in Section \ref{sec:Compound Outage Probability}, which includes its closed-form characterization in Theorem \ref{thm:PoutRx} (as one-dimensional convex optimization problem) and the worst channel distribution (which is a piece-wise constant scaling of the nominal distribution). Remarkably, the compound outage probability depends only on the nominal one and the relative entropy distance, all other details (e.g. nominal fading and noise distributions) being irrelevant. Properties of the compound outage probability are given in Propositions \ref{prop:P(d)}-\ref{prop:Pout(eps)}, and its two-regime asymptotic behavior is identified in Section \ref{sec:LimitingRegimes}. Specifically, as the nominal outage probability decreases (say by increasing the SNR), the compound outage probability first decreases linearly too, but after a certain threshold (equal to about the relative entropy distance when the latter is small), it decreases only logarithmically, i.e. very slowly, so that significant decrease is not possible anymore. Optimizing the transmit signal distribution in this regime does not bring in significant improvement either so that any reasonable distribution (e.g. isotropic signalling) will do as well. Compact, closed-from approximations are obtained for the compound outage probability in these two regimes using the tools of asymptotic analysis. Theorem \ref{thm:Pout*s} shows that the transmit signal distribution optimal for the nominal channel distribution is also optimal for the whole class, so that known optimal transmit covariance matrices (see e.g. \cite{Telatar}\cite{Goldsmith}-\cite{Jorcwieck}) can be "recycled".

Since relative entropy is not symmetric, Section \ref{sec:Class 2} investigates the impact of this asymmetry on the outage probability. Swapping the distributions (nominal and true) is shown to result in the error floor effect: the compound outage probability is bounded away from zero, does not matter how low the nominal outage (or how high the SNR) is. The error floor depends on the relative entropy distance: it decreases with it and when it is small, they are equal, so that the relative entropy distance also serves as the compound outage probability in this regime. Theorem \ref{thm:PoutRx2} provides a closed-from expression for the compound outage probability that depends on a unique solution of a single non-linear equation. An alternative characterization is via one-dimensional convex optimization. The worst-case channel distribution and properties of the compound outage probability are also given, including its two-regime behavior and compact approximations via asymptotic analysis.

Based on the results above, compound outage capacity is studied in Section \ref{sec:Compound Outage Capacity}. In Section \ref{sec:Lp distance}, the class of distributions is considered where the distance is defined via $L_p$ norm (a particular case of $p=2$ corresponds to popular mean-square-error estimation) and its outage probability is related to the results above. In particular, the compound outage with relative entropy distance serves as a lower bound for this case and the error floor effect is present as well. Section \ref{sec:Conclusion} concludes the paper. Proofs are given in the Appendix.

\section{System Model, Outage Probability and Capacity}
\label{sec:System model}

\begin{figure}[htbp]
\centerline{\includegraphics[width=2in]{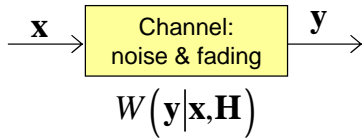}} \caption{Generic discrete-time basedband  MIMO block-fading channel model. No assumptions on noise and fading distributions are made.}
\label{fig:Figure_channel}
\end{figure}

Let us consider a generic discrete-time baseband multiple-input multiple-output channel as shown in Fig. \ref{fig:Figure_channel}, where $\bf{x}$ and $\bf{y}$ are the input (transmitted) and output (received) vectors (or sequences), and $\bf{H}$ denotes channel state. In the general case, the channel is described by the conditional probability distribution of $\bf{y}$ given $\bf{x}$ and $\bf{H}$, $W\left(\bf{y} |\bf{x},\bf{H}\right)$, and the mutual information (per channel use) supported by the channel for a given distribution of $\bf{x}$ and channel state $\bf{H}$ is $I(\bf{x};\bf{y}|\bf{H})$. We assume that the channel is block-fading (non-ergodic), i.e. a particular channel realization $\bf{H}$ is selected in the beginning and stays fixed for the whole duration of codeword transmission; next codeword will see a different channel realization \footnote{With a slight modification in notations, this block-fading model can also be extended to the case where each codeword sees a finite number of channel realizations, e.g. as in \cite{Caire}\cite{Varanasi}, and our results will hold in that case as well.}. Channel fading distribution is described by its probability density function $f(\mathbf{H})$. Most of our results will hold in this generic scenario, which includes as special cases frequency-selective (inter-symbol interference) or frequency-flat (no ISI) Gaussian MIMO channels\footnote{and can also be extended to include the finite block-length regime considered in \cite{Polyanskiy}.}.

We will not assume any particular fading and noise distribution (except for examples) so that our results are general and apply to \emph{any} such distribution. The transmitted signal, receiver noise and the channel are assumed to be independent of each other. We also assume that the transmitter does not known the channel but only has a partial knowledge of its distribution (as explained later on); channel knowledge at the receiver is irrelevant in the block-fading (i.e. quasi-static) environment.

In the special case of Gaussian MIMO channel, the channel model becomes
\begin{equation}
\label{channel}
\bf{y} = \bf{Hx}+ \boldsymbol \xi
\end{equation}
where $\boldsymbol \xi$ is the additive white complex circularly-symmetric Gaussian noise and $\bf{H}$ is the channel matrix, whose $(i,j)$-th entry is the channel gain between $i$-th output and $j$-th input (e.g receive/transmit antennas, time or frequency slots etc., see e.g. \cite{Caire} for more details). As is well-known, when the noise is Gaussian, the optimal signaling is also Gaussian (see e.g. \cite{Telatar}) so that complex circularly-symmetric Gaussian $\bf{x}$ is optimal. Its distribution is completely characterized by its covariance matrix $\bf{R_x} = {\mathbb E}[xx^\dagger]$, where $^\dagger$ denotes Hermitian conjugation. For a given channel realization $\bf{H}$, a given transmit covariance $\bf{R_x}$ achieves the celebrated log-det mutual information in this Gaussian MIMO channel \cite{Telatar}\cite{Foshini},
\begin{equation}
\label{I(x;y)}
I(\bf{x};\bf{y}|\bf{H}) = \ln |\bf{I} + \gamma \bf{H}\bf{R_x}\bf{H}^\dagger|
\end{equation}
where $|\cdot|$ denotes determinant, $\bf{I}$ is the identity matrix, $\gamma$ is the average SNR per Rx antenna.

Main performance metrics in the block-fading regime are outage probability and outage capacity \cite{Biglieri}\cite{Tse}\cite{Caire} \footnote{It can be further shown that the outage probability is the best achievable average codeword error probability \cite{Verdu}-\cite{Effros}.}.
Outage probability is the probability that the channel is not able to support the target rate $R$. When the transmitter knows the channel distribution (but not the channel itself), the outage probability is
\begin{equation}
\label{Pout}
P_{out}(R)= \underset{tr \mathbf{R_x} \leq P_T}{\min} \Pr \{I(\mathbf{x;y|H})< R \}
\end{equation}
where $\Pr \{I(\mathbf{x;y|H})< R \}$ is the outage probability for a given transmit covariance and the minimization is over all possible transmit covariance matrices satisfying the power constraint, $P_T$ is the total transmit power\footnote{In the general case, the minimization in \eqref{Pout} is over all possible distributions of $\bf{x}$ subject to the power constraint. To simplify notations, we will use below the notations of \eqref{Pout}, which apply to the Gaussian MIMO channel in \eqref{channel}, with understanding that the same results hold for the general MIMO channel as well.}. Outage capacity is defined as the largest possible rate such that the outage probability does not exceed the target value $\delta$,
\begin{equation}
\label{Cout}
C_{0\delta}= \max\{R: P_{out}(R) \le \delta \}
\end{equation}
Clearly, $P_{out}(C_{0\delta}) = \delta$. Finally, one may also consider the outage probability and capacity for a given (fixed) transmit covariance. Achievability of the outage capacity/probability follows from the compound channel capacity theorem \cite{Blackwell}-\cite{Lapidoth}\cite{Caire} (which guarantees an existence of a code that works on every channel in the no-outage set); see also \cite{Verdu}\cite{Effros} for a modern treatment using the concept of information density.

\section{Compound Outage Probability for a Class of Channels}
\label{sec:Compound Outage Probability}

Consider the scenario where the transmitter has only partial channel distribution information. Namely, it knows that the channel probability density function (PDF) $f(\bf{H})$ is within a certain distance of the nominal distribution $f_0(\bf{H})$. We use the relative entropy as a measure of the distance between two distributions, so that all feasible distributions $f$ satisfy the following inequality:
\begin{equation}
\label{fClass}
D(f||f_0) = \int f \ln \frac{f}{f_0}d\mathbf{H} \leq d
\end{equation}
where $D(f||f_0)$ is the relative entropy or Kullback-Leibler distance between the distributions, and $d$ is the maximum possible distance in the uncertainty set to which $f$ belongs. Throughout the paper we assume that $d<\infty$. In this scenario, the definition in \eqref{Pout} does not apply (since the true distribution $f$ is not known) but can be generalized to
\begin{equation}
\label{Pout*}
P_{out}^* = \underset{tr \mathbf{R_x} \leq P_T}{\min} \underset{D(f||f_0)\leq d}{\max}\Pr \{I(\mathbf{x;y|H})< R \}
\end{equation}
and the outage capacity can be defined as in \eqref{Cout} with the substitution $P_{out} \rightarrow P_{out}^*$. Its achievability also follows from the compound channel capacity theorem \cite{Root}\cite{Lapidoth}\cite{Caire} or from \cite{Verdu}\cite{Effros}, since the optimal signalling does not depend on the true channel distribution, but only on the nominal one (and also the relative entropy distance $d$). This problem setup models a practical situation where the channel distribution information is obtained from measurements or physical modeling, which are never perfect. It also accounts for the fact that the estimated channel distribution may change with time in dynamic scenarios. We term $P_{out}^*$ in \eqref{Pout*} "compound outage probability" since it is a performance measure of a class of channel distributions rather than a single distribution. This approach parallels the work on compound channel capacity \cite{Dobrushin}-\cite{Lapidoth}\cite{Wiesel}-\cite{Denic} where the channel is not known to the transmitter but it is known to belong to a certain class.

To characterize the compound outage probability $P_{out}^*$, we adopt a two-step approach: first, we characterize the outage probability for a given Tx covariance matrix (i.e. no minimization in \eqref{Pout*}, which also represent a practical situation where the transmit covariance is set a priory); then it is minimized over all feasible transmit covariances.

\subsection{Step 1: Compound Outage for a Given Tx Covariance}

When the Tx covariance $\mathbf{R_x}$ is given\footnote{In the general case, this corresponds to a given distribution of $\bf{x}$.}, the compound outage probability is
\begin{equation}
\label{PoutRx}
P_{out}(\mathbf{R_x}) = \underset{D(f||f_o)\leq d}{\max}\Pr \{I(\mathbf{x;y|H})< R \}
\end{equation}
Its characterization is strikingly simple in the generic scenario, i.e. for any noise and nominal fading distribution.

\vspace{5pt}
\begin{thm}
\label{thm:PoutRx}
For a given Tx covariance $\mathbf{R_x}$ and arbitrary nominal fading distribution $f_0$, the outage probability in \eqref{PoutRx} can be expressed as
\begin{equation}
\label{Pout_s}
P_{out}(\mathbf{R_x}) = \underset{s \geq 0}{\min} [s \ln (1 + (e^{1/s}-1)\varepsilon)+sd]
\end{equation}
where
\begin{equation}
\label{eps}
\varepsilon = \int\limits_{I(\mathbf{x;y|H})< R} f_0 d\mathbf{H}
\end{equation}
is the nominal outage probability (i.e. the outage probability under the nominal channel distribution). The worst channel distribution $f^*$ (the maximizer in \eqref{PoutRx}) is given by

\begin{eqnarray}
\label{f*} f^{*}= \frac{(e^{1/s^*}-1)\ell(\mathbf{H})+1}{(e^{1/s^*}-1)\varepsilon+1} f_0
\end{eqnarray}
where $s^*$ is the minimizing $s$ in \eqref{Pout_s}, and $\ell(\mathbf{H})$ is the indicator of the outage set: $\ell(\mathbf{H})=1$ if $I(\mathbf{x;y|H})< R$ and 0 otherwise.
\end{thm}
\begin{IEEEproof}
see Appendix.
\end{IEEEproof}
\vspace{5pt}

Note that Theorem \ref{thm:PoutRx} effectively reduces the infinite-dimensional optimization problem in \eqref{PoutRx} (the optimization there is over the set of all admissible distributions $f$) to one-dimensional convex optimization in \eqref{Pout_s}, which can be effectively solved using numerical algorithms. This is accomplished using Lagrange duality theory (see Appendix for details). As we will see below, this is not the only advantage: \eqref{Pout_s} also provides a number of insights unavailable from \eqref{PoutRx}. It is remarkable that the nominal outage distribution enters the compound outage probability in \eqref{Pout_s} only via the nominal outage probability $\varepsilon$, all other its details being irrelevant, i.e. two different nominal distributions with the same nominal outage probability will produce the same compound outage probability.

\begin{figure}[htbp]
\centerline{\includegraphics[width=3in]{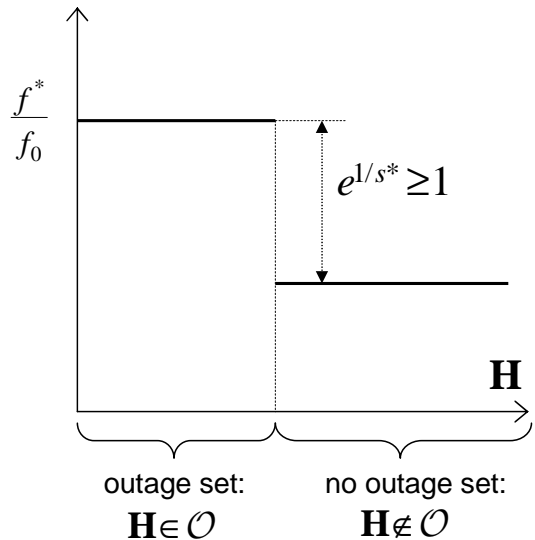}} \caption{The worst channel distribution over the nominal one. Note that the worst distribution is a piece-wise constant scaling of the nominal one.}
\label{fig:f*/f0}
\end{figure}

Note that the maximizing density $f^*$ in \eqref{f*} mimics the nominal one $f_0$ in a piece-wise constant manner:
\begin{eqnarray}
\label{f*a}
\frac{f^*}{f_0} = \left\{ \begin{array}{cc}
\frac{e^{1/s^*}}{(e^{1/s^*}-1)\varepsilon+1}, & \textrm{if $\mathbf{H}\in \mathcal{O}$} \\
\frac{1}{(e^{1/s^*}-1)\varepsilon+1} & \textrm{if $\mathbf{H}\notin \mathcal{O}$} \\\end{array}
\right.
\end{eqnarray}
where $\mathcal{O} = \{ \mathbf{H}: \ell(\mathbf{H})=1 \}$ is the outage set, so that the right-hand side of \eqref{f*a} is independent of $\mathbf{H}$ in each set and $f^*$ is a scaled up version of $f_0$ in the outage set and scaled down otherwise - see Fig.\ref{fig:f*/f0}.

An additional advantage of \eqref{Pout_s} is that the resulting optimization problem there is convex, i.e. the function
\begin{equation}
\label{L(s)}
L(s) = s \ln (1 + (e^{1/s}-1)\varepsilon)+sd
\end{equation}
is convex in $s>0$ (see Appendix) so that $P_{out}(\mathbf{R_x}) = \min_{s \geq 0} L(s)$ in \eqref{Pout_s} can be solved efficiently using any known numerical algorithm (the solution is unique and satisfies $dL(s)/ds = 0$). Alternatively, the tools of asymptotic analysis (see e.g. \cite{Efgrafov}\cite{Olver}) can be used to obtain approximations (see Section \ref{sec:LimitingRegimes}). This dual representation also provides a number of insights, as indicated below.

Let us now consider the outage probability in \eqref{Pout_s} as a function $P_{out}(d)$ of the distance $d$. A number of its properties follow.

\vspace{5pt}
\begin{prop}
\label{prop:P(d)}
For a given covariance $\mathbf{R_x}$, the compound outage probability $P_{out}(d)$  as a function of distance $d$ has the following properties:

1) $P_{out}(d)$ is concave in $d\geq0$.

2) $P_{out}(d=0)=\varepsilon$, i.e. the compound outage probability equals the nominal one when $d=0$.

3) $P_{out}(d)$ is a non-decreasing function of $d$, that is
\begin{eqnarray}
P_{out}(d_1)\leq P_{out}(d_2), \quad 0\leq d_1 < d_2<\infty.
\end{eqnarray}
and the equality holds if and only if $P_{out}(d_1)=P_{out}(d_2)=1, 0$ i.e. $P_{out}(d)$ is a strictly increasing function of the distance $d$ unless $P_{out}=1, 0$.
\end{prop}
\begin{IEEEproof}
see Appendix.
\end{IEEEproof}
\vspace{5pt}

\begin{prop}
\label{prop:Pout}
The compound outage probability $P_{out}$ has the following properties:

1) $P_{out}=1$ if and only if $\varepsilon=1$.

2) $P_{out}=0$ if and only if $\varepsilon=0$.

3) $P_{out}\geq\varepsilon$, and the equality holds if and only if $d=0$ or $\varepsilon=0, 1$.
\end{prop}
\begin{IEEEproof}
see Appendix.
\end{IEEEproof}
\vspace{5pt}

While in general the compound and nominal outage probabilities can be very different (as property 3 above shows), the compound outage takes on a limiting value (either 0 or 1) if and only if the nominal outage does so.

\begin{prop}
\label{prop:Pout(eps)}
The compound outage probability in \eqref{Pout_s} is a strictly-increasing, concave function of the nominal outage $\varepsilon$, i.e.
\begin{eqnarray}
P_{out}(\varepsilon_1) < P_{out}(\varepsilon_2), \quad 0\leq \varepsilon_1 < \varepsilon_2\leq 1.
\end{eqnarray}
with the boundary conditions $P_{out}(\varepsilon=0)=0$, $P_{out}(\varepsilon=1)=1$.
\end{prop}
\begin{IEEEproof}
see Appendix.
\end{IEEEproof}
\vspace{5pt}

\begin{prop}
\label{prop:Pout_bounds}
The compound outage probability $P_{out}$ in Theorem \ref{thm:PoutRx} can be bounded as follows:
\begin{equation}
\label{Pout_bounds}
\varepsilon \leq P_{out} \leq \min[d+(e-1)\varepsilon,1].
\end{equation}
\end{prop}
\begin{IEEEproof}
see Appendix.
\end{IEEEproof}
\vspace{5pt}

\subsection{Asymptotic Regimes}
\label{sec:LimitingRegimes}

We now consider the compound outage in \eqref{Pout_s} in two limiting regimes:

1) The uncertainty-dominated regime $\varepsilon \rightarrow 0$ and fixed $d$, i.e. the dominant source of outage events is from significant deviation of the true channel distribution from the nominal one (outage events under the nominal distribution can be neglected).

2) The nominal-outage dominated regime $d\rightarrow 0$ and fixed $\varepsilon$, i.e. when the impact of channel distribution uncertainty is negligible as outage events under the nominal distribution dominate the performance.

\vspace{5pt}
\begin{prop}
\label{prop:eps->0}
The outage probability $P_{out}$ in \eqref{Pout_s} in the low nominal outage regime, $\varepsilon \rightarrow 0$ and fixed $d>0$, is as follows:
\begin{equation}
\label{Pout eps->0} P_{out} = \frac{d}{\ln\frac{d}{\varepsilon}-\ln\ln\frac{d}{\varepsilon}}(1+o(1)).
\end{equation}
and the optimal (minimizing) $s^*$ in \eqref{Pout_s} is given by
\begin{equation}
\notag
 s^* = \frac{1}{\ln\frac{d}{\varepsilon}-\ln\ln\frac{d}{\varepsilon}}(1+o(1)).
\end{equation}
This is the the uncertainty-dominated regime (the main contribution to $P_{out}$ is coming from $d$ rather than $\varepsilon$).
\end{prop}
\begin{IEEEproof}
see Appendix.
\end{IEEEproof}
\vspace{5pt}

Further analysis shows that the approximations above (without $o(1)$ term) are accurate provided that $\varepsilon \ll d < 1$. Note from \eqref{Pout eps->0} that the main contribution to $P_{out}$ is coming from $d$ (i.e. the uncertainty) rather than $\varepsilon$ (i.e. the nominal outage) since $\ln(d/\varepsilon)$ is a slowly-varying function of $\varepsilon$, so that variations from the nominal channel distribution dominate the outage events. Also note that the relative entropy distance $d$ is directly related to the compound outage probability, which indicates that it is this distance that should be used as a measure of accuracy in estimating the channel distribution from measurements or physical modeling since it is directly related to the system performance (outage probability and capacity).

Let us now consider the nominal outage-dominated regime (i.e. fixed $\varepsilon$ and $d\rightarrow 0$).

\vspace{5pt}
\begin{prop}
\label{prop:d->0}
In the low channel distribution uncertainty regime, $d\rightarrow 0$ and fixed $\varepsilon$, the compound outage probability is
\begin{equation}
\label{Pout d->0}
P_{out} = \varepsilon + \sqrt{2d(1-\varepsilon)\varepsilon} + o(\sqrt{d})
\end{equation}
and the optimal $s^*$ is given by
\begin{equation} \notag
s^*=\sqrt{\varepsilon(1-\varepsilon)/(2d)}(1+o(1)),
\end{equation}

\end{prop}
\begin{IEEEproof}
see Appendix.
\end{IEEEproof}
\vspace{5pt}

Further analysis shows that the approximation above is accurate when $d \ll \varepsilon < 1$ and that the impact of uncertainty is negligible, $P_{out}\approx\varepsilon$, when $\sqrt d \ll \sqrt{\varepsilon}$. Comparing Proposition \ref{prop:eps->0} and Proposition \ref{prop:d->0}, one concludes that indeed there are two regimes in the behavior of $P_{out}(\varepsilon)$:

\begin{enumerate}
\item The uncertainty dominated regime (nominal outage is neglegible), when $\varepsilon \ll d < 1$ so that
\begin{equation}
\label{Pout eps<<d}
P_{out}\approx\frac{d}{\ln\frac{d}{\varepsilon}-\ln\ln\frac{d}{\varepsilon}}\sim \frac{d}{\ln(1/\varepsilon)}
\end{equation}
where $\sim$ means "scales as", so that $P_{out}$ depends linearly on $d$ but only logarithmically  (i.e. very slowly) on $\varepsilon$.
\item The nominal outage-dominated regime (uncertainty is negligible), when $d \ll \varepsilon < 1$ and
\begin{equation}
\label{Pout d<<eps}
P_{out}\approx \varepsilon + \sqrt{2d(1-\varepsilon)\varepsilon} \sim \varepsilon
\end{equation}
i.e. $d$ contributes very little to the outage probability.
\end{enumerate}

These two regimes immediately suggest some design guidelines related to the outage probability. In the uncertainty dominated regime, the main way to reduce outage probability is via decreasing the uncertainty of the channel distribution, e.g. via improved channel measurements or modeling; reducing the nominal outage probability is not efficient here, so that minimizing it via the optimal transmit covariance (or distribution in the general case) is not worth the effort - any reasonable covariance (e.g. isotropic signalling) will do as well. This approach, however, will bring little improvement in the nominal outage-dominated regime, where the only way to reduce the outage probability is via improving systems performance under the nominal fading, e.g. by increasing the SNR or optimizing the transmit covariance. Note that these conclusions hold for any nominal channel distribution (e.g. not limited to i.i.d. Rayleigh) and for any noise (not only Gaussian).

\begin{figure}[htbp]
\centerline{\includegraphics[width=3.5in]{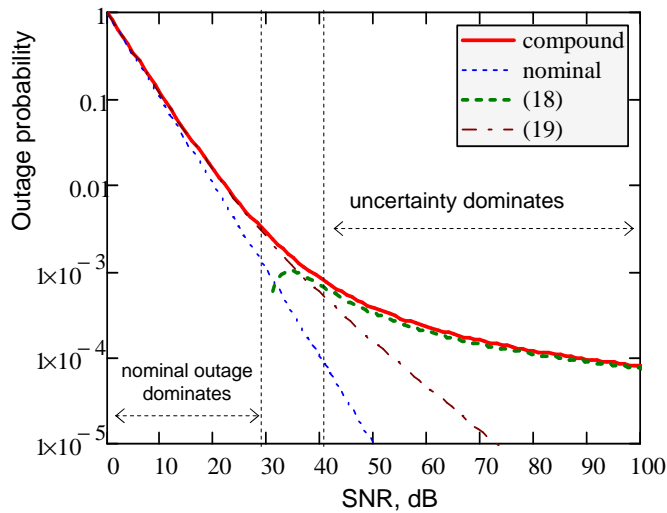}} \caption{Two-regime behavior of the compound outage probability. Its approximations in \eqref{Pout eps<<d} and \eqref{Pout d<<eps} and nominal outage $\varepsilon = 1/\mathrm{SNR}$ (set for convenience) are also shown; $d=10^{-3}$. Note that both approximations are accurate in their respective regimes. Decreasing the compound outage probability beyond about $10^{-3} (\approx d)$ requires exponentially high SNR and is not practical (it takes 60 dB extra to go from $10^{-3}$ to $10^{-4}$, while normally, i.e. without uncertainty, it would take only 10 dB).}
\label{fig:Pout.eps}
\end{figure}

As an example, consider a $1\times 1$ Rayleigh-fading channel with Gaussian noise, in which case $\varepsilon \sim 1/\gamma$, when $d \ll 1$ and $\gamma \gg 1$, so that
\setlength{\arraycolsep}{2pt}
\begin{eqnarray}
 P_{out}(\gamma)\sim \left\{ \begin{array}{cc} \frac{1}{\gamma} & \textrm{in regime 1} \ (\gamma \ll 1/d) \\
 \frac{d}{\ln \gamma} &\textrm{in regime 2} \ (\gamma \gg 1/d) \\ \end{array} \right.
\end{eqnarray}
i.e. the outage probability scales with SNR $\gamma$ as $1/\mathrm{SNR}$ in regime 1 but only as $1/\ln \mathrm{SNR}$ in regime 2, i.e. increasing the SNR is only efficient in the former case. When $d \ll 1$, increasing the SNR will first decrease $P_{out}$, but only down to about $d$ and after that point the decrease becomes logarithmically slow. From a practical perspective, it means that $P_{out}$ cannot be reduced significantly beyond $d$ by increasing the SNR. This observation indicates that the relative entropy distance $d$ is indeed an appropriate measure of channel distribution uncertainty in the non-ergodic (block-fading) mode. Fig. \ref{fig:Pout.eps} illustrates  the two-regime behavior of $P_{out}(\gamma)$.

This two-regime behavior can also be linked to the way channel distribution is obtained from measurements: a finite number of fading channel realizations are measured and the empirical channel distribution is derived based on it. However, the relative accuracy of this empirical distribution is always lower at the distribution tails, where fewer measurement points are available. On the other hand, when the average SNR is high, as in regime 2, an outage event takes place when the channel is very weak, i.e. at the distribution tail, so that the inaccuracy in the channel distribution estimation plays a dominant role there. Ultimately, low compound outage probability can only be achieved by insuring sufficiently high accuracy of the estimated distribution tail (small $d$), i.e. when a sufficient number of independent measurements fall into that region.

\subsection{Step 2: Minimizing over the Tx Covariance}

Using Theorem \ref{thm:PoutRx}, we are now in a position to characterize the compound outage probability in \eqref{Pout*}.

\vspace{5pt}
\begin{thm}
\label{thm:Pout*s}
Consider a class of fading channels in \eqref{fClass}. Its compound outage probability in \eqref{Pout*} can be found from
\begin{eqnarray}
\label{Pout*s}
P_{out}^* &=& \underset{s \geq 0}{\min}[s \ln (1 + (e^{1/s}-1)\varepsilon^*)+sd]
\end{eqnarray}
where $\varepsilon^* = \min_{tr \mathbf{R_x} \leq P_T} \varepsilon(\mathbf{R_x})$ is the optimized nominal outage probability, so that the outage-minimizing covariance for the class of channel distributions in \eqref{fClass} and for the nominal distribution $f_0$ are the same,
\begin{equation}
\label{argminPout(Rx)}
\arg \underset{tr \mathbf{R_x} \leq P_T}{\min}P_{out}(\mathbf{R_x}) = \arg \underset{tr \mathbf{R_x} \leq P_T}{\min} \varepsilon(\mathbf{R_x})
\end{equation}
\end{thm}

\begin{IEEEproof}
\begin{align}
\label{Pout*s} \notag
P_{out}^* &= \underset{tr \mathbf{R_x} \leq P_T}{\min} P_{out}(\mathbf{R_x})\\
\notag
&\overset{(a)}{=} \underset{tr \mathbf{R_x} \leq P_T}{\min} \underset{s \geq 0}{\min}[s \ln (1 + (e^{1/s}-1)\varepsilon)+sd]\\ \notag
&\overset{(b)}{=} \underset{s \geq 0}{\min}[s \ln (1 + (e^{1/s}-1)\varepsilon^*)+sd]
\end{align}
where (a) follows from \eqref{Pout_s} and (b) follows from the fact that $\min_{tr \mathbf{R_x} \leq P_T}$ and $\min_{s \geq 0}$ can be swaped and $\ln(\cdot)$ is a monotonic function, so that the minimization of the compound outage over $\mathbf{R_x}$ is equivalent to the minimization of the nominal outage.
\end{IEEEproof}
\vspace{5pt}

A significance of Theorem \ref{thm:Pout*s} is that the infinite-dimensional optimization in \eqref{Pout*} is reduced to a one-dimensional convex optimization in \eqref{Pout*s} and, furthermore, the optimal Tx covariance is the same as for the nominal channel, so that a significant number of known results \cite{Telatar}\cite{Goldsmith}-\cite{Jorcwieck}\cite{Caire}
apply directly to the compound fading channel as well, i.e. no new search of optimal covariances is required.

When the compound outage $P_{out}^* $ in \eqref{Pout*s} is considered as a function of the distance $d$, $P_{out}^*(d)$, its properties mimic those in Proposition \ref{prop:P(d)} with the substitution $\varepsilon \rightarrow \varepsilon^*$. Also, the results and conclusions of Section \ref{sec:LimitingRegimes} hold under this substitution. In particular, optimizing the transmit covariance (distribution) is worth the effort only in the nominal outage dominated regime.

\section{Compound Outage Probability for the $D(f_0||f)\leq d$ Class}
\label{sec:Class 2}

Since relative entropy is not symmetric, i.e. $D(f||f_0)\neq D(f_0||f)$, we consider in this section the constraint
\begin{equation}
\label{fClass 2}
D(f_o||f) = \int f_0 \ln \frac{f_0}{f}d\mathbf{H} \leq d
\end{equation}
to see the impact of the order on the obtained results. One property of the compound outage probability $P_{out}$ in \eqref{PoutRx} for this class of distributions is immediate.
\begin{prop}
\label{prop:Pout2bound}
For a given Tx covariance matrix, the compound outage in \eqref{PoutRx} under the distribution class in \eqref{fClass 2} is bounded as follows:
\begin{align}
\label{Pout2bound}
P_{out} &\ge 1 - e^{-d} + e^{-d}\varepsilon \\
\label{Pout2bound1}
&\ge 1 - e^{-d}
\end{align}
and the bounds are achievable. When $d\ll1$,
\begin{align}
\label{Pout2bound2}
P_{out} \ge d + \varepsilon
\end{align}
\end{prop}
\begin{IEEEproof}
\eqref{Pout2bound} follows from the fact that the lower bound is achievable by
\begin{align}
\label{f2bound} \notag
f(\mathbf{H}) = p \delta(\mathbf{H}) + (1-p)f_0(\mathbf{H})
\end{align}
where $p=1-e^{-d}$ and $\delta(\cdot)$ is the Dirac delta function, i.e. by the distribution that mimics the nominal one except for placing mass $p$ at zero (where the mutual information is zero). The bound in \eqref{Pout2bound1} is trivial and is achievable when $e^{d}-1 \gg \varepsilon$. \eqref{Pout2bound2} follows from $e^{-d} \approx 1-d$ for $d\ll1$.
\end{IEEEproof}
\vspace{5pt}

An important conclusion is immediate from \eqref{Pout2bound1}: $P_{out}(\varepsilon=0) \ge 1-e^{-d}$, i.e. there is a saturation (or error floor) effect in the behavior of $P_{out}(\varepsilon)$: even though $\varepsilon \rightarrow 0$ (e.g. by $\mathrm{SNR}\rightarrow \infty $), $P_{out} \nrightarrow 0$. From \eqref{Pout2bound2}, $P_{out} \ge d$ when $d\ll1$ in this regime, i.e. cannot be made smaller than the relative entropy distance $d$, does not matter how large the SNR (or how small the nominal outage) is. This is in contrast to \eqref{Pout eps->0}, where $P_{out} \rightarrow 0$ when $\varepsilon \rightarrow 0$, even though logarithmically slowly (i.e. no saturation). The absence of saturation in the latter case should not however be overestimated, since the convergence $P_{out} \rightarrow 0$ is logarithmically slow in $\varepsilon$, i.e. requires exponentially large SNR, so that for all practical purposes, $P_{out}$ also saturates around $d$, as was indicated in Section \ref{sec:LimitingRegimes}. Note also that \eqref{Pout2bound2} places $d$ and $\varepsilon$ on equal footing, re-enforcing our earlier conclusion that $d$ is an adequate measure of fading uncertainty in the non-ergodic regime.

Let us now characterize precisely the compound outage probability for the distribution class in \eqref{fClass 2}.

\vspace{5pt}
\begin{thm}
\label{thm:PoutRx2}
Consider the compound outage probability for the distribution class in \eqref{fClass 2} and a given Tx covariance $\mathbf{R_x}$,
\begin{equation}
\label{PoutRx2}
P_{out}(\mathbf{R_x}) = \underset{D(f_0||f)\leq d}{\max}\Pr \{I(\mathbf{x;y|H})< R \}
\end{equation}
It can be expressed in the following form:
\begin{equation}
\label{PoutRx2a}
P_{out}(\mathbf{R_x}) = \frac{\lambda^*}{\mu-1}\varepsilon
\end{equation}
where $\mu \geq 1$ is a unique solution of
\begin{equation}
\label{mu}
\frac{\mu^{\varepsilon}(\mu-1)^{1-\varepsilon}}{\mu-1+\varepsilon} = e^{-d}
\end{equation}
and $\lambda^* = e^{-d} (\mu-1)^{\varepsilon}{\mu}^{1-\varepsilon}$. The maximizing density $f^*$ in \eqref{PoutRx2} is given by
\begin{equation}
\label{f*2}
f^* = \frac{\lambda^* f_0}{\mu - \ell(\mathbf{H})}
\end{equation}
An alternative characterization of $P_{out}(\mathbf{R_x})$ is
\begin{equation}
\label{Pout_s2}
P_{out}(\mathbf{R_x}) = \underset{\lambda \geq 0}{\min} \left[\lambda \varepsilon \left(\frac{1}{\mu-1} + \ln \frac{\mu}{\mu-1}\right) + \lambda \left(d - \ln\frac{\mu}{\lambda}\right)\right]
\end{equation}
where
\begin{equation}
\label{mu2}
\mu = \frac{1}{2} \left(1 + \lambda + \sqrt{(1-\lambda)^2 + 4 \lambda \varepsilon}\right)
\end{equation}
\end{thm}
\begin{proof}
see Appendix.
\end{proof}

\begin{figure}[htbp]
\centerline{\includegraphics[width=3in]{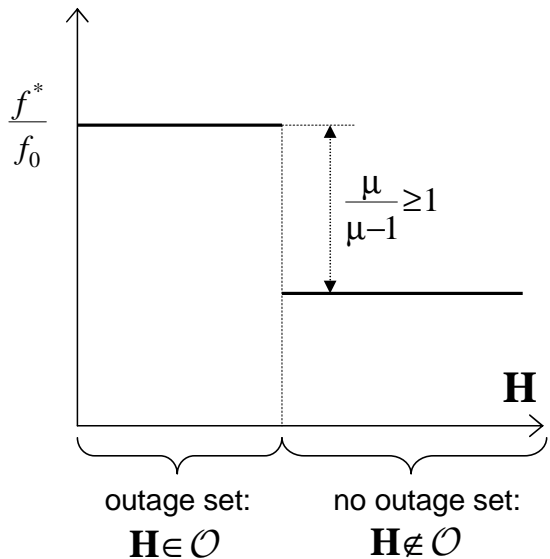}} \caption{The worst channel distribution over the nominal one. Note that the worst one is a piece-wise constant scaled version of the nominal one, which follows the same tendency as for the other uncertainty set.}
\label{fig:f*/f0-2}
\end{figure}

Note that Theorem \ref{thm:PoutRx2} reduces the infinite-dimensional optimization problem in \eqref{PoutRx2} to the closed-from solution in \eqref{PoutRx2a} that depends on a unique solution of a single monotonic non-linear equation in \eqref{mu} (can be efficiently found using any of known numerical techniques). The alternative representation in \eqref{Pout_s2} is a one-dimensional convex optimization problem. It follows from \eqref{f*2} that the maximizing density $f^*$ mimics the nominal one $f_0$, albeit with different constants in the outage and no outage sets,
\begin{eqnarray}
f^*=\left\{ \begin{array}{cc} \frac{\lambda^*}{\mu-1} f_0, & \textrm{if $\mathbf{H}\in \mathcal{O}$} \\ \frac{\lambda^*}{\mu} f_0 & \textrm{if $\mathbf{H}\notin \mathcal{O}$} \\\end{array}
\right.
\end{eqnarray}
where $\mathcal{O} = \{ \mathbf{H}: \ell(\mathbf{H})=1 \}$ is the outage set, so that $f^*$ is a scaled up version of $f_0$ in the outage set and scaled down otherwise\footnote{Note that $\lambda^* \le \mu \le 1+\lambda^*$, with 1st equality iff $\varepsilon=0$ and 2nd one iff $\varepsilon=1$.} (to keep the normalization fixed), which is what one would intuitively expect to maximize the outage probability. Fig. \ref{fig:f*/f0-2} illustrates this behavior.

Let us now consider asymptotic regimes.

\begin{prop}
\label{prop:Pout2eps->0}
The compound outage probability $P_{out}(\mathbf{R_x})$ in Theorem \ref{thm:PoutRx2} behaves in the low nominal outage regime $\varepsilon \rightarrow 0$ and fixed $d$ as follows:
\begin{equation}
\label{Pout2eps->0}
P_{out} = 1-e^{-d} + o(1).
\end{equation}
\end{prop}
\begin{proof}
see Appendix.
\end{proof}
\vspace{5pt}
Note that \eqref{Pout2eps->0} implies that $P_{out}(\varepsilon=0) = 1-e^{-d}$, i.e. the lower bound in \eqref{Pout2bound} is tight. It also re-affirms the saturation (error floor) effect discussed above. It is also interesting to note that the error floor $1-e^{-d}$ depends on the distance $d$ only, all other details (e.g. the nominal distribution $f_0$) being irrelevant. Further analysis shows that the approximation in \eqref{Pout2eps->0} is accurate, i.e. $P_{out} \approx 1-e^{-d}$, provided that $\varepsilon \ll e^d-1$. This is also consistent with the lower bound in \eqref{Pout2bound} (3rd term is negligible under the latter condition).

\begin{figure}[htbp]
\centerline{\includegraphics[width=3.5in]{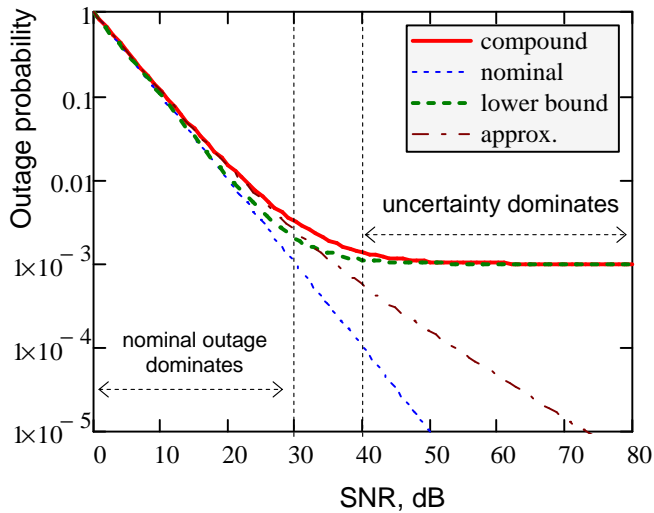}} \caption{Two-regime behavior of the compound outage probability and its lower bound in \eqref{Pout2bound}, the approximation in \eqref{Pout2d->0approx}, and the nominal outage $\varepsilon = 1/\mathrm{SNR}$; $d=10^{-3}$. Note the saturation (error floor) effect: the compound outage cannot be reduced below $10^{-3}(= d)$, in agreement with \eqref{Pout2bound1}, does not matter how high the SNR (or how low the nominal outage) is.}
\label{fig:Pout-2.eps}
\end{figure}

\begin{prop}
\label{prop:Pout2d->0}
The compound outage probability $P_{out}(\mathbf{R_x})$ in Theorem \ref{thm:PoutRx2} behaves in the small uncertainty regime $d\rightarrow 0$ and fixed $\varepsilon$ as follows:
\begin{equation}
\label{Pout2d->0}
P_{out} = \varepsilon+\sqrt{2\varepsilon(1-\varepsilon)d} + o\left(\sqrt{d}\right)
\end{equation}
\end{prop}
\begin{proof}
see Appendix.
\end{proof}
\vspace{5pt}
It is clear from \eqref{Pout2d->0} that $P_{out}(d=0)=\varepsilon$, so that the lower bound in \eqref{Pout2bound} is tight in this case. It is also obviously tight when $\varepsilon=1$ and, as we seen in Proposition \ref{prop:Pout2eps->0}, in the $\varepsilon=0$ case. In fact, numerical analysis shows that this lower bound is a good approximation of $P_{out}$ over the whole range of $\varepsilon$ - see Fig. \ref{fig:Pout-2.eps}. Comparing \eqref{Pout2d->0} to \eqref{Pout d->0}, we conclude that the compound outage probability is the same for the $D(f\|f_0)\le d$ and $D(f_0\|f)\le d$ uncertainty sets in the low uncertainty regime (i.e. the asymmetry of relative entropy does not have any effect here) while the same cannot be said about the low nominal outage regime (compare \eqref{Pout2eps->0} to \eqref{Pout eps->0}).

It can be shown (via numerical analysis - see Fig. \ref{fig:Pout-2.eps}) that the approximation in \eqref{Pout2d->0} is accurate, i.e.
\begin{equation}
\label{Pout2d->0approx}
P_{out} \approx  \varepsilon+\sqrt{2\varepsilon(1-\varepsilon)d}
\end{equation}
when $d<\varepsilon$ and that the effect of uncertainty is negligible, i.e. $P_{out} \approx \varepsilon$, when $d \ll \varepsilon$, which parallels the conclusions made in Section \ref{sec:LimitingRegimes}.

Fig. \ref{fig:Pout-2.eps} illustrates the two-regime behavior of the compound outage probability and validates the approximation above. It also demonstrates that the lower bound in \eqref{Pout2bound} is quite tight over the whole SNR (nominal outage) range and thus can be used instead of the true compound outage probability for design/analysis purposes to estimate the impact of channel distribution uncertainty.

We would like to emphasize that the results above hold for an arbitrary nominal channel distribution $f_0$ (e.g. Rayleigh, Rician, Nakagami, Log-Normal, correlated and/or non-identically distributed, etc.) - it enters into the compound outage probability only via the nominal outage $\varepsilon$, and also for arbitrary noise.

When the compound outage probability is minimized over transmit covariance (distribution) $\mathbf{R_x}$,
\begin{equation}
\label{Pout*2}
P_{out}^* = \underset{tr \mathbf{R_x} \leq P_T}{\min} \underset{D(f_0||f)\leq d}{\max}\Pr \{I(\mathbf{x;y|H})< R \}
\end{equation}
some of the results above (namely, \eqref{Pout2bound}-\eqref{Pout2bound2}, \eqref{Pout2eps->0}, and \eqref{Pout2d->0}, \eqref{Pout2d->0approx} when $\varepsilon<1/2$) apply directly via the substitution $\varepsilon \rightarrow \varepsilon^*$, i.e. the optimal transmit covariance is the same as for the nominal outage.  Furthermore, while optimizing the covariance to reduce the nominal outage also reduces the compound outage probability in the low uncertainty regime (i.e. in \eqref{Pout2d->0}), it has negligible impact in the uncertainty-dominated regime in \eqref{Pout2eps->0} so that any reasonable covariance (e.g. isotropic signalling) will work equally well (provided that $\varepsilon^* \ll e^d-1$).

\section{Compound Outage Capacity}
\label{sec:Compound Outage Capacity}
Let us now consider the compound outage capacity, which extends the definition of outage capacity in \eqref{Cout} to a class of distributions. Consider first the case when the transmit covariance is fixed:
\begin{equation}
\label{Ceps}
C_\delta= \max\{R: \max_f \Pr \{I(\mathbf{x;y|H})< R\} \le \delta \}
\end{equation}
where $f$ belongs to a distribution uncertainty class, $D(f||f_0)\leq d$ or $D(f_0||f)\leq d$.

It is clear that $C_\delta \leq C_{0\delta}$ in general, where the optimization over the transmit covariance is either used or not in both cases simultaneously. Using the compound outage probability results obtained above, the compound outage capacity can be characterized more precisely.

\begin{prop}
\label{prop:Ceps d->0}
The compound outage capacity in the low-uncertainty regime $d \ll \varepsilon <1$ for both uncertainty sets, $D(f||f_0)\leq d$ and $D(f_0||f)\leq d$, equals the nominal outage capacity,
\begin{align}
\label{Ceps d->0}
C_\delta \approx C_{0\delta}
\end{align}
\end{prop}
\begin{IEEEproof}
Follows directly from \eqref{Pout d<<eps} and \eqref{Pout2d->0approx}.
\end{IEEEproof}

\begin{prop}
\label{prop:Ceps2 bound}
The compound outage capacity for the $D(f_0||f)\leq d$ uncertainty set satisfies the following inequality,
\begin{align}
\label{Ceps2 bound}
C_\delta =0 \ \forall \delta < 1-e^{-d}
\end{align}
In particular, no transmission is possible at $\delta < d$: $C_\delta =0 \ \forall \delta < d$.
\end{prop}
\begin{IEEEproof}
Follows from \eqref{Pout2bound} and from $1-e^{-d} \leq d$.
\end{IEEEproof}

\begin{prop}
\label{prop:Ceps2 bound}
The compound outage capacity for the $D(f_0||f)\leq d$ uncertainty set in the $d \ll 1$ regime satisfies the following inequality,
\begin{align}
\notag
C_\delta \leq C_{0(\delta - d)}
\end{align}
i.e. $d$ serves as a measure of loss in the target outage probability due to the channel distribution uncertainty.
\end{prop}
\begin{IEEEproof}
Follows from \eqref{Pout2bound} and $1-e^{-d} + e^{-d}\delta \approx d + \delta$ for $d \ll 1$.
\end{IEEEproof}

When optimization over the transmit covariance (distribution) is done, the results above also apply with the substitution $\varepsilon \rightarrow \varepsilon^*$ and $C_{0\delta}$ meaning the optimized nominal outage capacity.

\section{Compound Outage Probability via $L_p$ Distance Constraint}
\label{sec:Lp distance}

Let us now consider the channel distribution uncertainty class of the form $\|f-f_0\|_p \leq d$, where $\|f\|_p = (\int |f|^p d\mathbf{H})^{1/p}$ is $L_p$ norm, $p \geq 1$. It can be characterized as follows.

\begin{prop}
\label{prop:P1 bound}
The compound outage probability for the $\|f-f_0\|_p \leq d$ uncertainty class, $p \geq 1$, can be bounded as follows:
\begin{align}
\label{Pp}
P_p &= \max_{\|f-f_0\|_p \leq d} \Pr \{I(\mathbf{x;y|H})< R\} \\
\label{P1}
& \geq P_1 \\
\label{P1 bound 1}
& \geq \max [P_{out}^1(d^2/2), P_{out}^2(d^2/2)] \\
\label{P1 bound 2}
& \geq 1 - e^{-d^2/2} + e^{-d^2/2}\varepsilon \\
\label{P1 bound 3}
& \ge  1 - e^{-d^2/2}
\end{align}
where $P_{out}^1(d)$ and $P_{out}^2(d)$ are the compound outage probabilities as functions of distance $d$ for the $D(f||f_0)\leq d$ and $D(f_0||f)\leq d$ classes.
\end{prop}
\begin{IEEEproof}
\eqref{P1} follows from the norm inequality $\|f-f_0\|_p \leq \|f-f_0\|_1$. \eqref{P1 bound 1} follows from Pinsker inequality $D(f||f_0) \geq \frac{1}{2} \|f-f_0\|_1^2$ \cite{Fedotov} and the fact that $\|f-f_0\|_1 = \|f_0-f\|_1$. \eqref{P1 bound 2} and \eqref{P1 bound 3} follow from \eqref{Pout2bound}.
\end{IEEEproof}

Note that there is an error floor effect here as well: $P_p(\varepsilon=0) \geq 1 - e^{-d^2/2} > 0$ $\forall d>0$. The $p=2$ case corresponds to the widely-used mean square error criterion (including channel estimation and measurements), so that there is an error floor for the MSE uncertainty class as well. Finally, when the compound outage probability is minimized over the transmit covariance (distribution), the same inequalities hold (where $P_{out}^{1(2)}$ and $\varepsilon$ are also minimized), and the error floor is not affected by this. When $d$ is small (so that the deviation of $f$ from $f_0$ is also small), the $L_p$ and relative entropy distances can be shown to have the same order of magnitude, so that the corresponding compound outage probabilities will scale similarly.

\section{Conclusion}
\label{sec:Conclusion}

Compound outage probability and capacity of a class of fading MIMO channels with partial channel distribution information have been introduced and studied. These concepts generalize well-known and widely used concepts of outage probability and capacity of fading channels with completely known distribution to the case where only partial knowledge of distribution is available. Relative entropy distance is used as a measure of uncertainty, which is shown to be related directly to the compound outage probability. Since relative entropy distance is not symmetric, two uncertainty classes are considered and worst-case fading distributions are identified for both. A number of properties, bounds and approximations of the compound outage probability are given. The transmit covariance matrix optimized for the nominal outage probability is shown to be also optimal for the compound one. The nominal fading distribution enters into the compound outage probability only via the nominal outage probability, all other details being irrelevant, i.e. two different nominal distributions having the same nominal outage will also produce the same compound outage probability. Behavior of the compound outage probability reveals two distinct regimes: uncertainty-dominated regime and nominal-outage dominated one. While increasing the average SNR or optimizing the transmit covariance to reduce the outage probability is effective for the latter, it has only negligible effect in the former, which immediately suggest a design alternative (via reducing uncertainty rather than increasing the SNR or optimizing the transmit covariance). All these results are very general as they hold for arbitrary nominal fading distribution and also for arbitrary noise (i.e. not only Gaussian).

\section{Acknowledgment}

The research leading to these results has received funding from the European Community's Seventh Framework Programme (FP7/2007-2013) under grant agreement no. INFSO-ICT-223844 and the Cyprus Research Promotion Foundation under the grant ARTEMIS.

\section{Appendix}
\label{sec:Appendix}

\subsection{Proof of Theorem \ref{thm:PoutRx}}
The problem in \eqref{PoutRx} can be presented in this form \footnote{An additional constraint $f\ge0$ is not included since our solution explicitly satisfies it.}:
\begin{eqnarray}
\label{Pout_Opt}
\max_f \int \ell(\mathbf{H}) f d\mathbf{H}, \textrm{ s.t. } \int f \ln \frac{f}{f_0} d\mathbf{H} \leq d, \int f d\mathbf{H} = 1.
\end{eqnarray}
The problem is clearly convex (since the objective is linear and the constraint set is convex). Furthermore, strong duality holds (i.e. the duality gap is zero), since Slater condition (see e.g. \cite{Boyd}) is satisfied. Therefore, the KKT conditions are sufficient for optimality \cite{Boyd}. The Lagrangian of this problem is
\begin{eqnarray}
\label{Lagrangian}
L = \int \ell(\mathbf{H}) f d\mathbf{H} - s \left(\int f \ln \frac{f}{f_0} d\mathbf{H} - d \right) - \mu \left(\int f d\mathbf{H} - 1 \right),
\end{eqnarray}
and, taking the variational derivative (see e.g. \cite{Gelfand}) of $L$ with respect to $f$, one obtains the KKT conditions:
\begin{align}
\label{KKT conditions}
\ell(\mathbf{H}) - s \left(\ln \frac{f}{f_0} + 1\right) - \mu = 0 \\
\label{KKT mu}
\int f d\mathbf{H} - 1 = 0 \\
\label{KKT slackness}
s \left(\int f \ln \frac{f}{f_0} d\mathbf{H} - d \right) = 0 \\
\label{KKT s >=0}
s \ge 0
\end{align}
It is straightforward to see that the complementary slackness condition \eqref{KKT slackness} implies that the second term is zero, i.e. the optimum is achieved on the boundary. Using \eqref{KKT conditions} and \eqref{KKT mu}, one obtains, after some manipulations,
\begin{eqnarray}
\label{f* Proof}
f^* = f_0 \frac{e^{\ell(\mathbf{H})/s^*}}{1+\varepsilon (e^{1/s^*}-1)}
\end{eqnarray}
where $f^*$ and $s^*$ are the solutions of \eqref{KKT conditions}-\eqref{KKT s >=0}, from which \eqref{f*} follows. To obtain \eqref{Pout_s}, note that, due to zero duality gap, the solution to \eqref{Pout_Opt} equals to that of the dual problem,
\begin{eqnarray}
\label{Pout dual}
P_{out}(\mathbf{R_x}) = \min_{s \ge 0} L(s)
\end{eqnarray}
where $L(s)$ is the dual function,
\begin{align}
\label{L(s) proof}
\notag
L(s) &= \max_f \{L\} \text{\ s.t. \eqref{KKT mu}} \\
 &= s \ln (1 + (e^{1/s}-1)\varepsilon)+sd
\end{align}
and $\mu$ was eliminated using \eqref{KKT mu}. Since $L(s)$ is the dual function, it is convex. Furthermore, when $0 < \varepsilon < 1$,
\begin{align}
\notag
\frac{d^2 L(s)}{ds^2} = \frac{\varepsilon (1-\varepsilon)e^{1/s}}{s^2(1+(e^{1/s}-1)\varepsilon)^2} > 0
\end{align}
for $0<s<\infty$, i.e. strictly convex, so that the problem in \eqref{Pout dual} has a unique solution. Since $P_{out} = \varepsilon$ when $\varepsilon = 1,0$ (see Proposition \ref{prop:Pout}), the problem in \eqref{Pout dual} has always a unique solution, which can be found from $dL(s)/ds = 0$ using any known numerical algorithm. The tools of asymptotic analysis (e.g. \cite{Efgrafov}\cite{Olver}) allow one to obtain a number of approximations.

\subsection{Proof of Proposition \ref{prop:P(d)}}
\label{Proof of prop P(d)}
1) To prove item 1, note that $P_{out}$ in \eqref{Pout_s} is a point-wise minimum of a set of affine functions of $d$ (indexed by $s$) and therefore is concave (see e.g. \cite{Boyd}).

2) Using Lyapounov inequality,
\begin{align}
\notag
\left(\mathbb{E}|x|^p\right)^{1/p} \leq \left(\mathbb{E}|x|^q \right)^{1/q}
\end{align}
where $0 < p \le q$ and $\mathbb{E}$ denotes expectation, for random variable $x=e^{\ell(\mathbf{H})}$ and $p=1/s_1, q=1/s_2$, $0 \leq s_2\leq s_1$, one obtains
\begin{align}
\notag
\left(\mathbb{E}e^{\ell(\mathbf{H})/s_1}\right)^{s_1} \leq \left(\mathbb{E}e^{\ell(\mathbf{H})/s_2}\right)^{s_2}
\end{align}
and taking $\log$,
\begin{align}
\notag
s_1 \ln \mathbb{E}e^{\ell(\mathbf{H})/s_1} \leq s_2 \ln \mathbb{E}e^{\ell(\mathbf{H})/s_2}
\end{align}
so that $F(s) = s \ln \mathbb{E}e^{\ell(\mathbf{H})/s} = s \ln (1+(e^{1/s}-1)\varepsilon)$ is a non-increasing function of $s$. Therefore,
\begin{align}
\label{Pout(d=0)=F(infinity)}
P_{out}(d=0) = \min_{s \geq 0} F(s) = \lim_{s \rightarrow \infty} F(s) = \varepsilon
\end{align}

3) Define $L_i(s) = s \ln (1 + (e^{1/s}-1)\varepsilon)+sd_i$, $i=1,2$, $d_1 < d_2$. Then, $L_1(s) \leq L_2(s)$ with equality iff $s=0$, in which case $L_1(s)=L_2(s)=1$ or 0 (if $\varepsilon>0$ or = 0). Taking $\min_{s \geq 0}$ of both sides, one obtains $P_{out}(d_1) \leq P_{out}(d_2)$ with equality iff $P_{out}(d_1) = P_{out}(d_2) = 1, 0$.

\subsection{Proof of Proposition \ref{prop:Pout}}

\label{Proof of prop Pout)}

1) Using \eqref{f* Proof},
\begin{eqnarray}
\label{Pout_s*_Proof}
P_{out} = \frac{e^{1/s^*}\varepsilon}{1+\varepsilon (e^{1/s^*}-1)}
\end{eqnarray}
from which it follows that $P_{out}=1$ iff $\varepsilon=1$.

2) Item 2 is proved in the same way as item 1 above.

3) The inequality is due to that fact that the distribution uncertainty set always includes the nominal distribution. The "if" part follows from items 1 and 2 above and from item 2 of Proposition \ref{prop:P(d)}. The "only if" part is verified using \eqref{Pout_s*_Proof}.

\subsection{Proof of Proposition \ref{prop:Pout(eps)}}

\label{Proof of prop Pout(eps))}
It mimics the proof of item 3 in Proposition \ref{prop:P(d)}. Define $L_i(s) = s \ln (1 + (e^{1/s}-1)\varepsilon_i)+sd$, $i=1,2$, $\varepsilon_1 < \varepsilon_2$. Then, $L_1(s) \leq L_2(s)$ with equality iff $s=0$, in which case $L_1(s)=L_2(s)=1$. Taking $\min_{s \geq 0}$ of both sides, one obtains $P_{out}(\varepsilon_1) \leq P_{out}(\varepsilon_2)$ with equality iff $P_{out}(\varepsilon_1) = P_{out}(\varepsilon_2)= 1$, which is possible only if $\varepsilon_1 = \varepsilon_2= 1$. The concavity of $P_{out}(\varepsilon)$ follows from the fact that it is a point-wise minimum of a set of concave functions of $\varepsilon$ (indexed by $s$), see \eqref{Pout dual}.

\subsection{Proof of Proposition \ref{prop:Pout_bounds}}

\label{Proof of prop:Pout_bounds}
The lower bound was proved above. The upper bound follows from the following:
\begin{align}
\notag
P_{out} &= \min_{s \geq 0} [s \ln (1 + (e^{1/s}-1)\varepsilon)+sd]\\ \notag
&\leq \ln (1 + (e-1)\varepsilon)+d \\ \notag
&\leq (e-1)\varepsilon+d
\end{align}
where 1st inequality is obtained by setting $s=1$ and 2nd one follows from $\ln (1+x) \leq x$ for $x\geq0$.

\subsection{Proof of Proposition \ref{prop:eps->0}}

\label{Proof of prop:eps->0}
Since $L(s)$ in \eqref{L(s)} is convex, a unique minimum in \eqref{Pout_s} can be found by setting the derivative to zero, $dL(s)/ds=0$, which can be expressed as
\begin{align}
\label{L(s)'=0}
d + \ln(1+y) = \frac{y+\varepsilon}{1+y}\ln\left(1+\frac{y}{\varepsilon}\right)
\end{align}
where $y=\varepsilon(e^{1/s}-1)$. It is  straightforward to see that $y\rightarrow 0$ as $\varepsilon \rightarrow 0$ so that \eqref{L(s)'=0} can be transformed to
\begin{align}
\label{L(s)'=0 2}
\frac{d}{\varepsilon}+o\left(\frac{d}{\varepsilon}\right) = \left(1+ \frac{y}{\varepsilon}\right) \ln \left(1+\frac{y}{\varepsilon}\right)
\end{align}
where we have used $\ln(1+y)=y+o(y)$ and $1/(1+y)=1-y+o(y)$. Using the techniques of asymptotic analysis (see e.g. \cite{Efgrafov}\cite{Olver}), the following equation
\begin{align}
%\label{}
x \ln x = u
\end{align}
has the solution  when $u\rightarrow \infty$,
\begin{align}
%\label{}
x = \frac{u}{\ln u - \ln \ln u} \left(1+o\left(\frac{1}{\ln u}\right)\right)
\end{align}
Setting $x=1+y/\varepsilon=e^{1/s}$, $u=(1+o(1))d/\varepsilon$, one obtains, after some lengthy but straightforward manipulations, the solution $y^*$ of \eqref{L(s)'=0 2} and corresponding $s^*$:
\begin{align}
\notag
y^* &= \frac{d}{\ln \frac{d}{\varepsilon} - \ln \ln \frac{d}{\varepsilon}} (1+o(1)) \\
\notag
s^* &= \frac{1}{\ln \frac{d}{\varepsilon} - \ln \ln \frac{d}{\varepsilon}} (1+o(1))
\end{align}
so that, after some further manipulations,
\begin{align}
\label{Pout eps->0 Proof}
P_{out} = L(s^*) = \frac{d}{\ln \frac{d}{\varepsilon} - \ln \ln \frac{d}{\varepsilon}} (1+o(1))
\end{align}

\subsection{Proof of Proposition \ref{prop:d->0}}

\label{Proof of prop:d->0}
In the $d\rightarrow 0$ and fixed $\varepsilon$ regime, the minimizer $s^*$ in \eqref{Pout_s} is $s^*\rightarrow \infty$ (as follows from \eqref{Pout(d=0)=F(infinity)}), so that $y^*/\varepsilon = e^{1/s^*}-1 \rightarrow 0$ and \eqref{L(s)'=0} can be transformed to
\begin{align}
\notag
\frac{1}{2}\left(\frac{y}{\varepsilon}\right)^2(1-\varepsilon)+o(y^2)= \frac{d}{\varepsilon}
\end{align}
from which one obtains
\begin{align}
\notag
y^* = \sqrt{\frac{2\varepsilon d}{1-\varepsilon}}(1+o(1))
\end{align}
so that
\begin{align}
\notag
s^* = \sqrt{\frac{\varepsilon (1-\varepsilon)}{2d}}(1+o(1))
\end{align}
and
\begin{equation}
\label{Pout d->0 proof}
P_{out} = \varepsilon + \sqrt{2d(1-\varepsilon)\varepsilon} + o(\sqrt{d})
\end{equation}
Further analysis shows that the approximation above (without $o(\sqrt{d})$ term) is accurate when $d \ll \varepsilon <1$.

\subsection{Proof of Theorem \ref{thm:PoutRx2}}

This proof follows along the same lines as that of Theorem \ref{thm:PoutRx}, which is summarized below. The optimization problem is:
\begin{eqnarray}
\label{Pout_Opt2}
\max_f \int \ell(\mathbf{H}) f d\mathbf{H}, \textrm{ s.t. } \int f_0 \ln \frac{f_0}{f} d\mathbf{H} \leq d, \int f d\mathbf{H} = 1.
\end{eqnarray}
The problem is also convex and strong duality holds, so that the KKT conditions are sufficient for optimality. The Lagrangian is
\begin{eqnarray}
\label{Lagrangian2}
L = \int \ell(\mathbf{H}) f d\mathbf{H} - \lambda \left(\int f_0 \ln \frac{f_0}{f} d\mathbf{H} - d \right) - \mu \left(\int f d\mathbf{H} - 1 \right),
\end{eqnarray}
and the KKT conditions are:
\begin{align}
\label{KKT conditions2}
\ell(\mathbf{H}) + \lambda \frac{f_0}{f} - \mu = 0 \\
\label{KKT mu2}
\int f d\mathbf{H} - 1 = 0 \\
\label{KKT slackness2}
\lambda \left(\int f_0 \ln \frac{f_0}{f} d\mathbf{H} - d \right) = 0 \\
\label{KKT s >=02}
\lambda \ge 0
\end{align}
After some manipulations, one obtains
\begin{eqnarray}
\label{f* Proof 2}
f^* = \frac{\lambda^* f_0} {\mu - \ell(\mathbf{H})}
\end{eqnarray}
where $f^*$ and $\lambda^*$ are the solutions of \eqref{KKT conditions2}-\eqref{KKT s >=02}. $\lambda^*$ is found from \eqref{KKT slackness2}
\begin{eqnarray}
\label{lambda*2}
\lambda^* = e^{-d} (\mu-1)^{\varepsilon} \mu^{1-\varepsilon}
\end{eqnarray}
and $\mu$ is found from \eqref{KKT mu2}
\begin{eqnarray}
\label{mu*2}
\frac{\lambda^*\varepsilon}{\mu-1} + \frac{\lambda^*(1-\varepsilon)}{\mu} = 1
\end{eqnarray}
which can be transformed, after some manipulations, to \eqref{mu}.

The alternative characterization of $P_{out}(\bf{Rx})$ follows from the dual problem,
\begin{eqnarray}
\label{Pout dual 2}
P_{out}(\mathbf{R_x}) = \min_{\lambda \ge 0} L(\lambda)
\end{eqnarray}
where $L(\lambda)$ is the dual function,
\begin{align}
\label{L(lambda)2}
\notag
L(\lambda) &= \min_f \{L\} \text{\ s.t. \eqref{KKT mu2}} \\
&= \lambda\varepsilon\left(\frac{1}{\mu-1} + \ln\frac{\mu}{\mu-1}\right) + \lambda\left(d - \ln\frac{\mu}{\lambda}\right)
\end{align}
and $\mu$ is found from \eqref{mu*2},
\begin{equation}
\label{mu2proof}
\mu = \frac{1}{2} \left(1 + \lambda + \sqrt{(1-\lambda)^2 + 4 \lambda \varepsilon}\right).
\end{equation}

\subsection{Proof of Proposition \ref{prop:Pout2eps->0}}

\label{Proof prop:Pout2eps->0}
It is straightforward to see that the left-hand side of \eqref{mu} is a strictly increasing function of $\mu$, $\forall \mu \ge 0$, $\forall \varepsilon \in (0,1)$, which takes values in $[0,1]$, so that a solution is unique. It also follows that $\mu \rightarrow 1$ as $\varepsilon \rightarrow 0$ so that expanding $\mu(\varepsilon)$ in Taylor series, $\mu = 1+ a_1\varepsilon + o(\varepsilon)$, one obtains from \eqref{mu}
\begin{equation}
\ln \frac{a_1}{1+a_1} + o(1) = -d
\end{equation}
where we have used $\ln\mu = a_1 \varepsilon + o(\varepsilon)$, from which it follows that
\begin{equation}
\label{mu2eps->0}
\mu = 1 + \frac{\varepsilon}{e^d - 1} + o(\varepsilon)
\end{equation}
and also $\lambda = e^{-d}(1+o(1))$, so that
\begin{equation}
\label{Pout2eps->0Proof}
P_{out} = \frac{\lambda^*}{\mu-1}\varepsilon = 1 - e^{-d} + o(1)
\end{equation}
\eqref{mu2eps->0} suggests that the approximation in \eqref{Pout2eps->0Proof} (without $o(1)$ term) is accurate when $\varepsilon \ll e^d - 1$, which is also confirmed by numerical analysis.

\subsection{Proof of Proposition \ref{prop:Pout2d->0}}

\label{Proof prop:Pout2d->0}
It follows from \eqref{mu} that $\mu \rightarrow\infty$ as $d\rightarrow 0$ so we use the expansion
\begin{equation}
\ln \left(1 - \frac{1}{\mu}\right) = - \frac{1}{\mu} + \frac{1}{2\mu^2} + o\left(\frac{1}{\mu^2}\right)
\end{equation}
to transform \eqref{mu} to
\begin{equation}
\frac{\varepsilon(1-\varepsilon)}{2\mu^2} + o\left(\frac{1}{\mu^2}\right) = d
\end{equation}
from which it follows that
\begin{equation}
\mu = \sqrt{\frac{\varepsilon(1-\varepsilon)}{2d}}(1 + o(1))
\end{equation}
and also that $\lambda^* = \mu (1+o(1))$. Using these, one finally obtains
\begin{equation}
P_{out} = \frac{\lambda^*}{\mu-1}\varepsilon = \varepsilon+\sqrt{2\varepsilon(1-\varepsilon)d} + o(\sqrt{d})
\end{equation}

\end{document}